\documentclass{aa}

\usepackage[varg]{txfonts}

\usepackage{color}
\usepackage{graphicx}
\usepackage{natbib}
\usepackage{pdflscape}

\begin{document}

\titlerunning{A VLBI survey of compact broad absorption line quasars with balnicity index BI$>$0}

\title{A VLBI survey of compact broad absorption line quasars with balnicity index BI$>$0}

\author{M. Kunert-Bajraszewska
     \and M. Ceg\l{}owski
     \and K. Katarzy\'nski
     \and C. Roskowi\'nski}

     \offprints{M. Kunert-Bajraszewska, \email{magda@astro.uni.torun.pl}}

         \institute{Toru\'n Centre for Astronomy, Faculty of Physics, Astronomy and Informatics, NCU, Grudziadzka 5, 87-100 Toru\'n, Poland}

\abstract{}
{Outflows manifest as broad absorption lines in the quasars spectra. Although outflows are one of the most common astrophysical processes in the Universe, the broad absorption line (BAL) quasars are rare. Radio emission is another tool that can help to understand the phenomenon of BAL quasars. The aim of this paper is to study their orientation and age by very long baseline interferometry (VLBI)
 imaging and radio-loudness parameter distribution.}
{
We performed high resolution radio observations of a new sample of ten BAL quasars using both the Very Long Baseline Array (VLBA) and the European VLBI Network (EVN) at 5\,GHz. All the selected sources have balnicity  indices (BI) more than 0 and radio flux densities less than 80 mJy at 1.4\,GHz. They are very compact with linear sizes of the order of a few tens of parsecs and radio luminosities at 1.4\,GHz above the FR\,I-FR\,II luminosity threshold.}
{

Most of the observed objects have been resolved at 5\,GHz showing one-sided, probably core-jet structures, typical for quasars. We discuss in detail their age and orientation based on the radio observations.
We then used the largest available sample of BAL quasars to study the relationships between the radio and optical properties in these objects.
We found that (1) the strongest absorption (high values of the balnicity index BI) is connected with the lower values of the radio-loudness parameter, log\,$\rm R_{I}<1.5$, and thus probably with large viewing angles; (2)  the large 
span of the BI values in each bin of the radio-loudness parameter indicates that the orientation is
only one of the factors influencing the measured absorption;  (3) most of the radio-loud BAL quasars are compact, low luminosity objects with a wide range of jet power (although the highest values of BI seem to be associated with the lower values of jet power).   In addition, we suggest that the short lifetime postulated for some compact AGNs could also explain  the scarcity of the large-scale radio sources among BAL quasars. 
}
{}


\maketitle


\section{Introduction}

Broad absorption troughs of high (C\,IV 1549 $\AA$ ) and low ionization resonant lines (Mg\,II 2800 $\AA$) 
are seen in the spectra of both the radio-quiet and radio-loud quasars. 
Sources showing these features are called broad absorption line (BAL) quasars and their fraction among the
 whole quasar population varies
from 15\,\% to 26\,\% depending on the definition used \citep{hewett2003, trump2006, knigge2008, gibson2009}.
BAL  quasars
(BALQSOs) are probably caused by the outflow of gas with high velocities and are part of the accretion process \citep{Weymann1991}. 
It is common to say that there are two models that could explain the presence of BALs in a small fraction of quasars, namely the orientation
and evolution schemes. However, the growing number of observations of the BALQSOs indicate the complexity of BAL phenomenon and make 
the nature and origin of BALQSOs still an open issue. Thus, selecting one of the two mentioned models would be a simplification.


\begin{table*}
\caption[]{Radio-loud BAL quasar sample.}
\label{basic}
\centering
\begin{tabular}{cccccccccccccc}
\hline

RA(J2000) & Dec(J2000) &
\multicolumn{1}{c}{\it z}&
\multicolumn{1}{c}{BI}&
\multicolumn{1}{c}{${\rm S_{1.4\,GHz}}$}&
\multicolumn{1}{c}{log${\rm L_{1.4\,GHz}}$}&
\multicolumn{1}{c}{${\rm S_{4.9\,GHz}}$}&
\multicolumn{1}{c}{${\rm S_{8.4\,GHz}}$}&
\multicolumn{1}{c}{ $\alpha_{fit}$}&
\multicolumn{1}{c}{$\rm log\,R_{I}$}&
\multicolumn{1}{c}{${\rm log\,P_{jet}}$}\\

h~m~s & $\degr$~$\arcmin$~$\arcsec$ & 
\multicolumn{1}{c}{}&
\multicolumn{1}{c}{($\rm km~s^{-1}$)} &
\multicolumn{1}{c}{(mJy)}&
\multicolumn{1}{c}{($\rm W~Hz^{-1}$)} & 
\multicolumn{1}{c}{(mJy)}&
\multicolumn{1}{c}{(mJy)}&
\multicolumn{1}{c}{}&
\multicolumn{1}{c}{}&
\multicolumn{1}{c}{($\rm W~Hz^{-1}$)}\\

(1)& (2)& (3) &(4)&   
\multicolumn{1}{c}{(5)}&
\multicolumn{1}{c}{(6)}& 
\multicolumn{1}{c}{(7)}&
\multicolumn{1}{c}{(8)}&  
\multicolumn{1}{c}{(9)}&
\multicolumn{1}{c}{(10)}&
\multicolumn{1}{c}{(11)}\\

\hline
08      11      02.931  &       50      07      24.52   &       1.84    &       371     &       24.9    &       26.8    &       10.9&11.9&0.41& 2.1     & 38.7 \\
08      42      24.395  &       06      31      16.78   &       2.46    &       1974&   51.0    &       27.4    &       25.6&21.2&0.49& 2.8     & 39.3 \\
08      56      41.566  &       42      42      53.94   &       3.06    &       295     &       20.0    &       27.2    &       20.5&13.7&0.19& 1.8     & 39.2 \\
10      40      59.802  &       05      55      24.78   &       2.44    &       4290&   42.2    &       27.3    &       8.1     &3.8&1.34&      2.5     & 39.2 \\
13      33      25.080  &       47      29      35.36   &       2.62    &       152     &       44.7    &       27.4    &       $-$     & $-$&$-$&        2.1     & 39.3 \\
14      01      26.163  &       52      08      34.63   &       2.97    &       70      &       37.1    &       27.4    &       $-$     & $-$&$-$&        2.0     & 39.4 \\
14      57      56.263  &       57      44      46.90   &       2.13    &       17      &       78.9    &       27.4    &       $-$     & $-$&$-$&        2.7     & 39.3 \\
21      07      57.683  &       -06     20      10.49   &       0.65    &       559     &       20.4    &       25.5    &       $-$     & $-$&$-$&        1.4     & 37.6 \\
22      38      43.578  &       00      16      48.05   &       3.47    &       13      &       36.9    &       27.6    &       $-$     & $-$&$-$&        2.2     & 39.6 \\
22      48      00.717  &       -09     07      44.93   &       2.11    &       16      &       36.7    &       27.1    &       $-$     & $-$&$-$&        2.3     & 39.0 \\
\hline
\end{tabular}
\begin{minipage}{173mm} 
\vspace{0.2cm}
{Description of the columns:
(1) and (2) source coordinates (J2000) extracted from FIRST, (3) redshift as measured from the SDSS, (4) balnicity index taken from \citet{trump2006}, 
(5) total flux density at 1.4\,GHz extracted from FIRST, (6) log of the radio luminosity at 1.4\,GHz, (7) and (8) total flux density at 4.9\,GHz and 8.4\,GHz taken
from \citet{dipompeo2011}, (9) spectral index using linear fit to all available radio data taken from \citet{dipompeo2011}
(10) radio-loudness, the radio-to-optical (i-band) ratio of the quasar core \citep{kimball2011b}, which were calculated
from {\it z}, ${\rm S_{1.4\,GHz}}$, $\rm M_{i}$ taken from \citet{trump2006}, and the assumption of a radio core spectral index of 0 and
an optical spectral index of -0.5, (11) jet power at 1.4\,GHz calculated using formula from \citet{sikora2013} (see also section \ref{balnicity}).}
\end{minipage}
\end{table*}

The spectroscopic and variability studies trying to characterize the UV absorber in BALQSOs generally indicate a complex behaviour of wind
outflows. BALs in radio-quiet and radio-loud quasars show wide range of depth changes from small to complete disappearance of troughs \citep{filiz2012}. This can be caused, for example, by a change in the location of absorber along the line of sight \citep{capellupo2013}. Recent observations of BAL variability in the sample of radio-loud BALQSOs
supports the orientation dependence of the observed outflow \citep{welling2014}.

Generally the orientation scenario implies that outflows are present in every quasar and the viewing angle of the source determines its detection.   
Thus, different wind geometries are proposed: near the equatorial plane \citep{Murray1995}, at mid inclination angles \citep{elvis, fine}, or very 
close to the jet axis \citep{ghosh2007, reynolds2013}. The radio variability study of some BALQSOs imply the existence of polar outflows
from the inner regions of a thin disk \citep{zhou2006, ghosh2007}. Nevertheless, objects with very high radio luminosities seem to be a minority among
BALQSOs \citep{shankar2008}. 

On the other hand the VLA FIRST survey showed that most of the radio-loud BALQSOs are very compact \citep{becker2000}.
The high resolution VLBI observations of BALQSOs performed so far \citep{jiang2003, kunert2007, doi2009, kun2010a, liu2008, monte2008, gawron2011, bruni2012, 
bruni2013, hayashi2013} have shown  that many of them are unresolved even on a parsec scale.
The analysis of the spectral shape, variability, and polarization properties of some of them
\citep{monte2008, kun2010a, liu2008} indicate that they are similar to compact steep spectrum
(CSS) objects and gigahertz peaked spectrum (GPS) objects, which are thought to be the progenitors of large-scale AGNs \citep{fanti95,odea98}.
Compact BALQSOs are not oriented along a particular line of sight, although they are more often observed farther from the jet axis than are normal quasars \citep{dipompeo2011, bruni2012}.
The alternative evolution scenario suggests then that every quasar has a BAL phase at the beginning of its lifetime \citep{becker2000, gregg2006}. 
The radio jets of the expanding source interact with the outflows and destroy the BALs before the source reaches the `adult' stage.

This paper is one of the two presenting high resolution radio observations and statistical analysis of compact radio-loud BALQSOs
selected from the most recent available catalogues of BALQSOs created by \citet{gibson2009} and \citet{trump2006}. The parent samples
were selected from the  \citet{trump2006} catalogue and VLA FIRST survey \citep{White1997} for further VLBI observations in two categories based on the values of absorption
index (AI) and balnicity index (BI), which classify the source as a BALQSO. It should be noted here that the BAL definition is diffuse nowadays.
The traditional BALs, quantified by the balnicity index (BI), are defined as having $\rm C_{IV}$ absorption troughs at least $\rm 2000\,km\,s^{-1}$ wide 
\citep{Weymann1991}. However, this could potentially exclude the so-called mini-BALs with BI=0. Therefore, the more liberal AI index, including quasars
with weaker and much narrower absorption features (within $\rm 3000\,km\,s^{-1}$) was introduced by \citet{trump2006}. 
The optical and radio analysis of BALQSOs from both groups (AI$>$0 \& BI=0 and BI$>$0)
revealed many differences between them indicating that they constitute two independent classes \citep{shankar2008, knigge2008}. However, if we
assume that the absorption by outflows comes from the continuum of velocity widths \citep{ganguly2007} then 
the study of both classes of objects is important in order to understand the origin and nature of BALs. 
 
In this paper we report high resolution radio observations of a sample of so-called traditional BALQSOs with 
balnicity index values BI$>$0 and thus low radio fluxes. The results of the study of mini-BALQSOs (AI$>$0 \& BI=0) are presented in \citet{ceg2015}.

\section{Sample selection}
\label{sample}

\subsection{Radio observations}
\label{radio_obs}
In order to build a sample of compact radio-loud BALQSOs we  matched the optical 
positions of BALQSOs from the \citet{trump2006} catalogue to FIRST coordinates \citep{White1997}
in a radius of 10 arcseconds. In practice it appeared that for most of the objects ($\sim 96\%$) the separation between FIRST and SDSS position is below 2 arcseconds.
We also limited our sample to sources with
integrated flux densities $S_{1.4\,{\rm GHz}}>2\,{\rm mJy}$ and side lobe probabilities
less than 0.1. 
In the next step we  excluded from this initial sample
objects with additional radio counterparts within 60 arcseconds of the SDSS position. These extended sources constituted $\sim 12\%$ of the initial sample. This approach allowed us to avoid doubts about  all the components belonging  to a single object and the identification of the core.
Finally, our sample consisted of 309 quasars classified as BALQSOs according to the
extended absorption index (AI) definition proposed by \citet{trump2006}, which includes BAL absorption
features at lower outflow velocities (within $\rm 3000\,km\,s^{-1}$). The traditional BALs are defined as  having 
$\rm C_{IV}$ absorption troughs at least $\rm 2000\,km\,s^{-1}$ wide and at least 10\% below the continuum at
maximum depth \citep{Weymann1991} and are quantified by the balnicity index (BI). \citet{trump2006} calculate
values of both the AI and BI indices for all quasars. Therefore, our sample in a natural way falls into two groups:
204 objects with AI$>$0 and BI=0 (AI sample) and 105 objects with AI$>$0 and BI$>$0, the so called `true BALQSOs' (BI sample). 

We have selected 31 sources with the largest 1.4\,GHz flux densities from both groups for further
high resolution VLBI observations. Since statistically the BI sources are weaker than AI quasars, this selection criterion means that the AI group (AI$>$0, BI=0) consists of the 16 bright sources
with flux densities $S_{1.4\,{\rm GHz}}>150\,{\rm mJy}$. The second BI group (AI$>$0, BI$>$0)
consists of 15 much weaker objects with flux densities in the range $\rm 20 - 80\,mJy$ at 1.4\,GHz. However, five of the weak sources from the BI sample
were already included in the sample of \citet{monte2008} and thus were not
observed by us. In this paper we report high resolution 5\,GHz observations of ten
quasars from the BI group, which are presented in Table~\ref{basic}.

All the C-band observations were made in February 2013 (EVN and VLBA) and December 2013 (VLBA) in 
 snapshot mode with phase referencing. Each target source together with its associated phase reference
source was observed for $\sim$60\,min including telescope drive times. In the case of VLBA
observations the correlation was performed with the new Distributed FX (DiFX) software correlator 
\citep{deller2011} at the National Radio Astronomy Observatory (NRAO) in Socorro. 
The European VLBI Network (EVN) observations were made using the antennas in Jodrell Bank, Westerbork,
Effelsberg, Onsala, Medicina, Torun, Shanghai, Noto, Yebes, and Hartebeesthoek.
The data were correlated at the Joint Institute for VLBI in Europe (JIVE) correlator in Dwingeloo.

Data reduction (including editing, amplitude calibration, instrumental phase corrections, and fringe-fitting)
was performed with the standard procedure using the NRAO AIPS\footnote{http://www.aips.nrao.edu} software.
After this stage the calibrated data were imported into Difmap package \citep{Shepherd1997} to produce 
the final Stokes {\it I} images and fit the sources with a number of discrete circular Gaussian components.
The fitting was done directly on the final, self-calibrated visibility data using the MODELFIT programme.
The final images of the radio-loud BALQSOs are presented in Fig.~\ref{images} and the modelfit parameters are listed in Table~\ref{observations}. The unresolved sources were classified as `Single' (S) and for the others we proposed `Core-Jet' (CJ) classification. We note here that we call this classification `suggested' because it is based only on observations made at one frequency. In the case of resolved structure the brightest component is referred to as C1.

\subsection{Statistical studies}
The statistical analysis of the properties of compact radio-loud BALQSOs presented in this paper was made using the 
most recent available catalogue of BALQSOs created by \citet{gibson2009} 
which is drawn from the Sloan Digital Sky Survey (SDSS) Data Release 5 instead of the 
\citet{trump2006} catalogue based on the SDSS/DR3. We then matched the optical positions from the \citet{gibson2009} catalogue to the FIRST radio positions with the same
selection criteria as described in section \ref{radio_obs} and found 303 radio-loud BALQSOs of which again 
$\sim 12\%$ are extended objects. Thus, the number of compact sources is 267.
It should be noted here that \citet{gibson2009} identify
the BALQSOs based on the `traditional' balnicity index BI \citep{Weymann1991} and a modified $\rm BI_{0}$ which integrates absorption starting from 
$\rm 0~km~s^{-1}$ instead of the traditional $\rm 3000~km~s^{-1}$ in the lines of $\rm C_{IV}$, $\rm Si_{IV}$, $\rm Al_{III}$, and $\rm Mg_{II}$. The group
of 267 compact radio-loud BALQSOs contains objects with $\rm BI_{0}>0$ in at least one of these lines.

We note here that the different methods used by \citet{gibson2009} and \citet{trump2006} in classifying objects as BALQSOs cause some differences between these
two catalogues. Two out of ten sources (1333+4729 and 1457+5744) selected from \citet{trump2006} and observed by us in the VLBI technique are not included in 
the \citet{gibson2009} catalogue.

Finally, we  note again that we removed from the sample studied here  about 12\% of the radio-loud BALQSOs with more than one radio counterpart in the radius of 60 arcseconds. We argue, that this operation did not introduce any additional selection effects to the obtained results. 
The calculated $\rm BI_{0}$ for most of the large-scale  objects is much less than 3000 and thus their removal did not affect the discussed conclusions, e.g. BI vs. log\,$\rm R_{I}$ relation or BI vs. jet power plot (see Figures~\ref{counts} and \ref{jet}).

Throughout the paper, we assume a cosmology with
${\rm
H_0}$=70${\rm\,km\,s^{-1}\,Mpc^{-1}}$, $\Omega_{M}$=0.3,
$\Omega_{\Lambda}$=0.7. The adopted convention for the spectral index
definition is $S\propto\nu^{-\alpha}$.

\section{Results and discussion}

\subsection{Characteristics of the VLBI sample}

\begin{figure*}
\centering
\includegraphics[width=16cm, height=5.5cm]{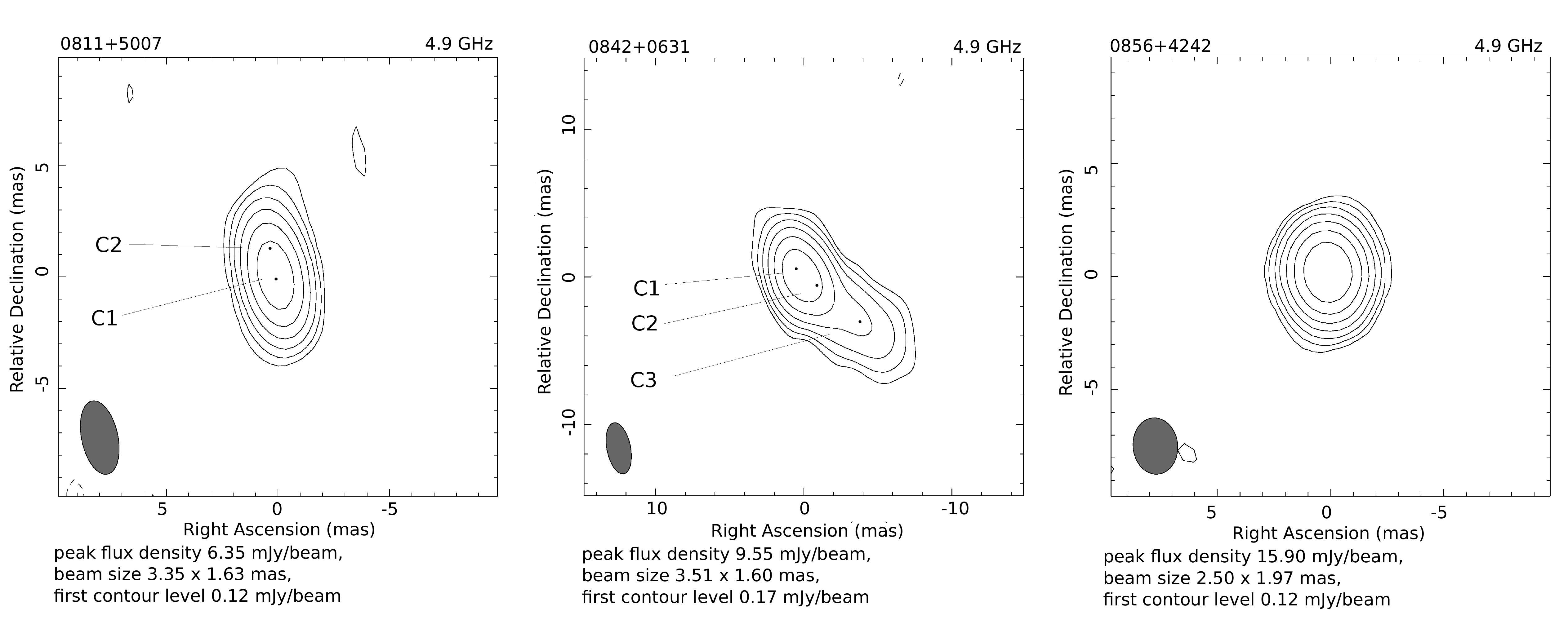} 
\includegraphics[width=16cm, height=6cm]{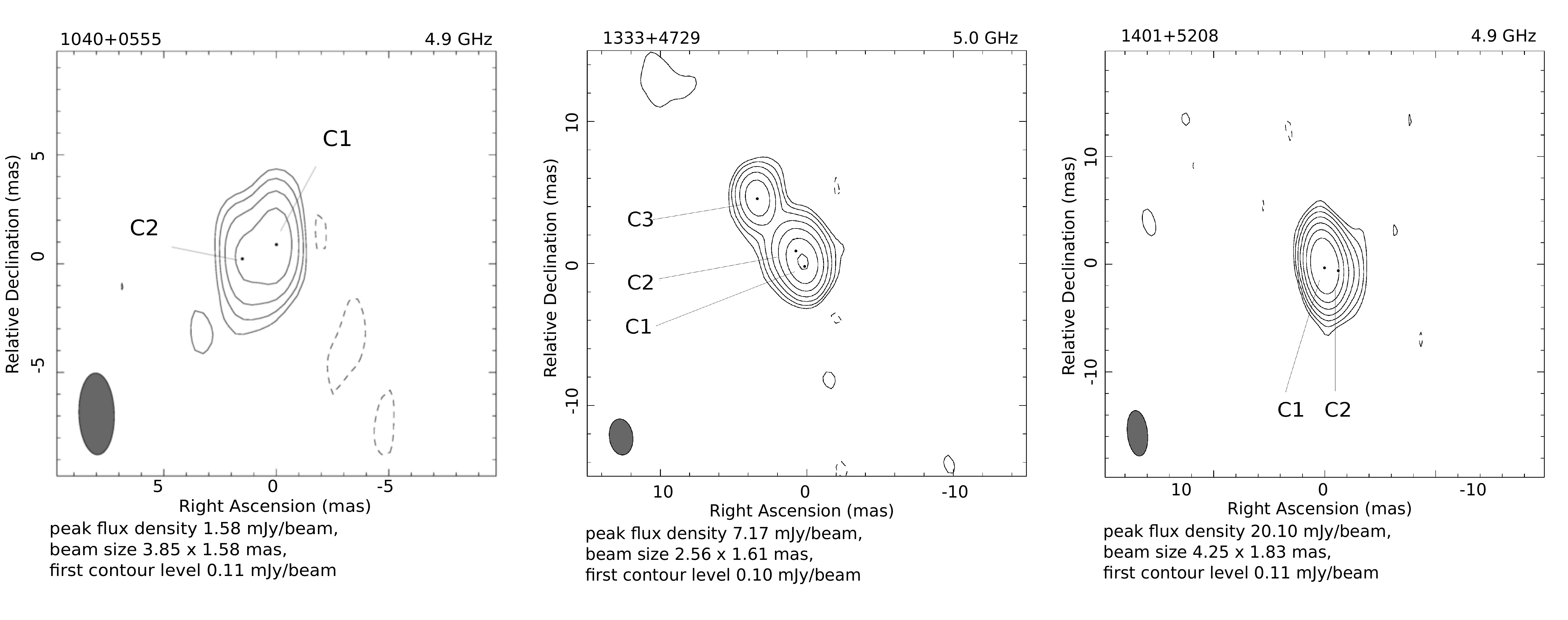}  
\includegraphics[width=16cm, height=6cm]{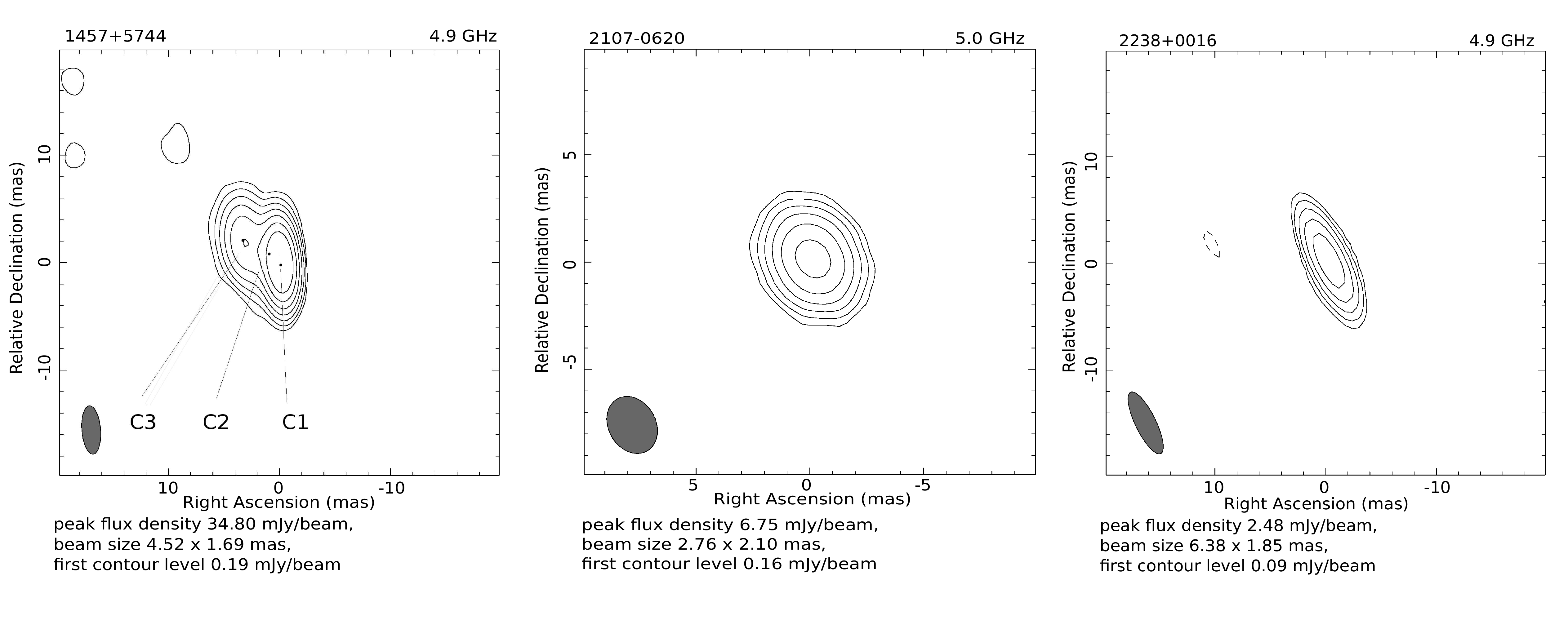} 
\includegraphics[width=5.5cm, height=5.5cm]{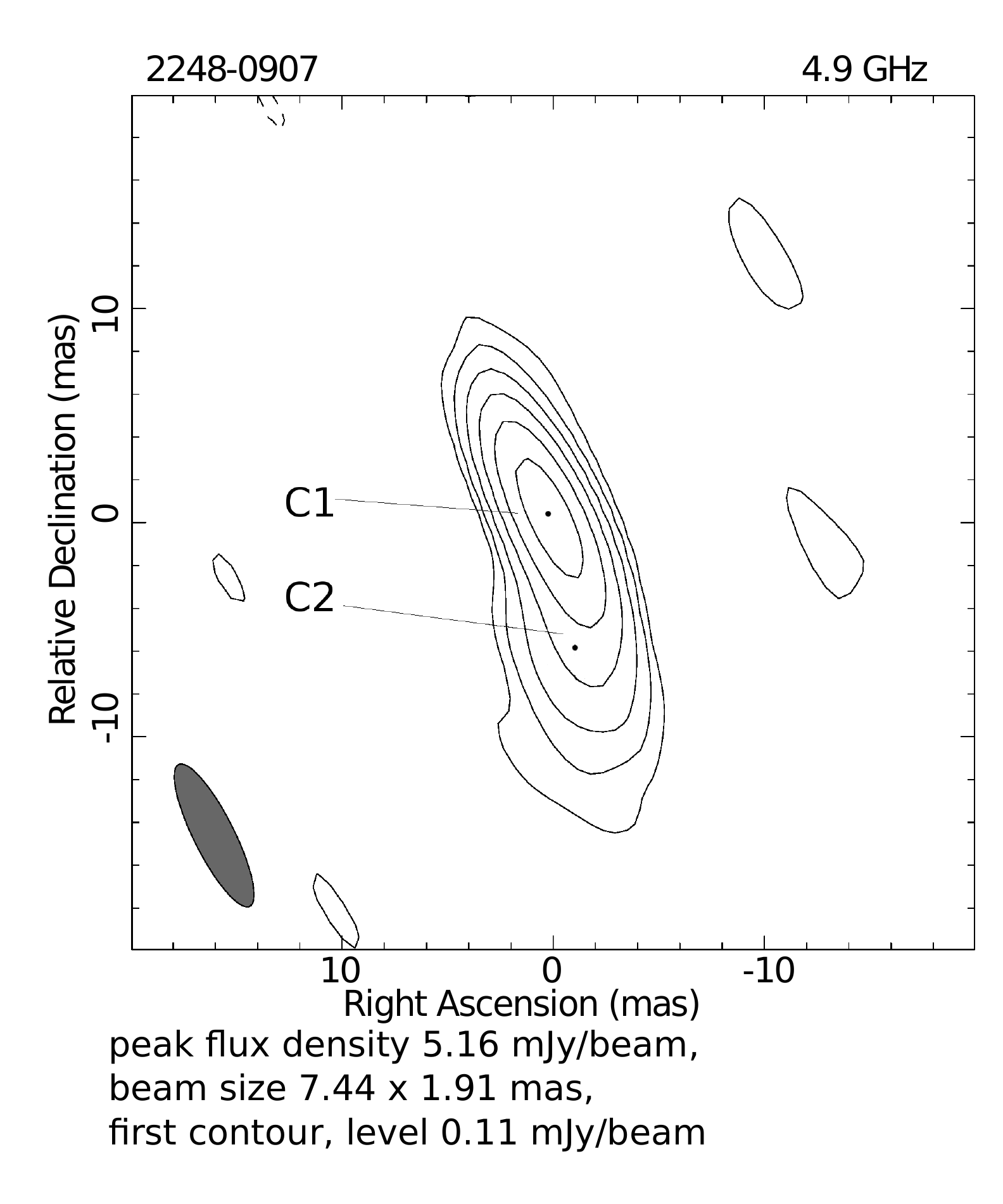}

\caption{Images of ten BALQSOs at 5\,GHz. Two  were made with EVN (1333+4729, 2107-0620) and the
others with VLBA. Contours increase by a factor of 2, and the first contour level corresponds to $\approx
3\sigma$.}
\label{images}
\end{figure*}

We have made high resolution 5\,GHz EVN/VLBA observations of  
ten radio-loud BALQSOs with balnicity index BI$>$0 and flux densities 
$S_{1.4\,{\rm GHz}}>20\,{\rm mJy}$. We detected all the objects, but three
of them (0856+4242, 2107-0620, 2238+0016) still remain unresolved at 5\,GHz. The resolved ones show one-sided
morphology probably indicating  the core-jet type of structure (Fig.~\ref{images}), typical for quasars. The small
linear sizes ($\rm \ll 1\,kpc$) of the observed quasars and their high luminosity (Table~\ref{sample}) may indicate
the young age of these AGNs. However, lack of observations at other radio frequencies
of most of the sources prevented us from analysing their spectra. Four of our objects have been 
observed with EVLA at 4.9 and 8.4\,GHz \citep{dipompeo2011} and two sources have X-ray flux estimations 
\citep{wang2008}. Simple linear models have been fitted to the spectra of four our sources by \citet{dipompeo2011} 
revealing a flat spectral index ($\rm \alpha < 0.5$) in the cases of 0811+5007 and 0856+4242, and a steep spectral index for resolved sources 0842+0631 and 1040+0555. 
The evolutionary status of the compact BALQSOs have already been discussed  by a few authors \citep{kunert2007,monte2008,
kun2010a,bruni2013}, who suggest that they belong to the class of young AGNs, namely the compact steep spectrum (CSS) sources (see also the 
discussion in section \ref{balnicity}).


\begin{table*}
\caption[]{Results of VLBA and EVN observations.}
\label{observations}
\centering
\begin{tabular}{ccccccccccccc}
\hline

RA(J2000) & Dec(J2000) &
\multicolumn{1}{c}{Comp.}&
\multicolumn{1}{c}{${\rm S_{5\,GHz}}$}&
\multicolumn{1}{c}{$\theta$}&
\multicolumn{1}{c}{LAS}&
\multicolumn{1}{c}{LLS}&
\multicolumn{1}{c}{${\rm T_{b}(vlbi)}$}&
\multicolumn{1}{c}{Type}\\

h~m~s & $\degr$~$\arcmin$~$\arcsec$ & 
\multicolumn{1}{c}{}&
\multicolumn{1}{c}{(mJy)}&
\multicolumn{1}{c}{(mas)}&
\multicolumn{1}{c}{(mas)}&
\multicolumn{1}{c}{(pc)}&
\multicolumn{1}{c}{($\rm 10^{9}~\delta^{-1}~K$)}&
\multicolumn{1}{c}{}\\
(1)& (2)& (3) &(4)&   
\multicolumn{1}{c}{(5)}&
\multicolumn{1}{c}{(6)}&
\multicolumn{1}{c}{(7)}&
\multicolumn{1}{c}{(8)}&
\multicolumn{1}{c}{(9)}\\
\hline
08      11      02.931  &       50      07      24.52   &       C1      &       6.4     &       0.7     & 1.5& 12.6&      2.9& CJ \\
                        &                               &       C2      &       0.8     &       0.6     & &               \\
08      42      24.395  &       06      31      16.78   &       C1      &       11.5    &       1.5     & 6.8& 55.1&      1.3& CJ \\
                        &                               &       C2      &       6.2     &       1.4     & &               \\
                        &                               &       C3      &       4.4     &       2.9     & &               \\
08      56      41.566  &       42      42      53.94   &       C       &       17.2    &       0.5     & 0.5& 3.8&       18.2& S \\
10      40      59.802  &       05      55      24.78   &       C1      &       1.8     &       0.8     & 2.1& 17.0&0.7& CJ               \\
                        &                                       &       C2      &       0.7     &       0.1     & &               \\
13      33      25.080  &       47      29      35.36   &       C1      &       5.7     &       0.6     & 6.0& 47.9&      4.0& CJ \\
                        &                               &       C2      &       5.1     &       1.7     & &               \\
                        &                               &       C3      &       3.3     &       0.5     & &               \\
14      01      26.163  &       52      08      34.63   &       C1      &       21.4    &       0.8     & 1.2& 9.3&       9.6& CJ \\
                        &                               &       C2      &       2.4     &       1.9     & &               \\
14      57      56.263  &       57      44      46.90   &       C1      &       33.5    &       0.4     & 4.2& 34.9&      40.1& CJ\\
                        &                               &       C2      &       7.8     &       1.6     & &               \\
                        &                               &       C3      &       7.2     &       1.9     & &               \\
21      07      57.683  &       -06     20      10.49   &       C       &       6.7     &       0.4     & 0.4& 2.8&       4.8& S  \\
22      38      43.578  &       00      16      48.05   &       C       &       2.9     &       1.2     & 1.2& 8.8&       0.6& S  \\
22      48      00.717  &       -09     07      44.93   &       C1      &       6.6     &       1.5     & 6.0& 49.9&      0.6& CJ \\
                        &                               &       C2      &       2.0     &       3.1     & &               \\

\hline
\end{tabular}
\begin{minipage}{137mm}
\vspace{0.2cm}
{Description of the columns:
(1) and (2) source coordinates (J2000) extracted from FIRST, (3) components as indicated on the images, (4) flux density measured with the VLBA or EVN,
(5) deconvolved major axis of the Gaussian fit, (6) largest angular size (LAS) measured at 5\,GHz image, LAS is defined as a separation between the two outermost Gaussian components or the size of the deconvolved component major axis (in the case of unresolved objects), (7) largest linear size (LLS) calculated based on the LAS, 
(8) brightness temperature calculated using equation \ref{vlbi}, (9) suggested type of radio morphology.
}
\end{minipage}
\end{table*}

Using the high resolution VLBI observations of our quasars we have calculated the brightness temperatures (equation \ref{vlbi} in the Appendix)
of their central components as indicated in the images and in  Table~\ref{observations}. The calculated values of ${\rm T_{b}(vlbi)}$ are in the
range $\rm 10^{8} - 10^{10}$, although these values should be treated as the
lower limits since we are restricted by the resolution of our observations. The brightness temperatures $\rm T_{b, 1.4\,GHz}\la 10^5\,K$ and radio luminosity  
$\rm L_{1.4\,GHz} \la 10^{24}\,W\,Hz^{-1}$ are thought to be associated with stellar components of luminous starburts \citep{condon1992,condon2013}. Therefore, the radio luminosities
and values of the brightness temperatures of our BALQSOs indicate the AGN origin of their radio emission. On the other hand their values of the
brightness temperatures do not exceed the inverse Compton brightness temperature limit of $\rm 10^{12}\,K$ and the empirical value of 
$\rm 10^{11}\,K$ \citep{lahteenmaki1999} that could indicate the emission boosting.  
However, three sources from our sample were previously reported
in the literature as polar BALQSOs \citep{zhou2006, ghosh2007}, which means objects with lines of sight close to the radio-jet axis. This classification
was made based on the high values of the variability brightness temperature ${\rm T_{b}(var)}$ of these quasars exceeding the theoretical 
limit of $\rm 10^{12}\,K$.
We  discuss this issue in the next paragraph.

\subsection{Statistics on the parent sample}
\subsubsection{Core radio-to-optical ratios}

\begin{figure}
\centering
\includegraphics[width=\columnwidth]{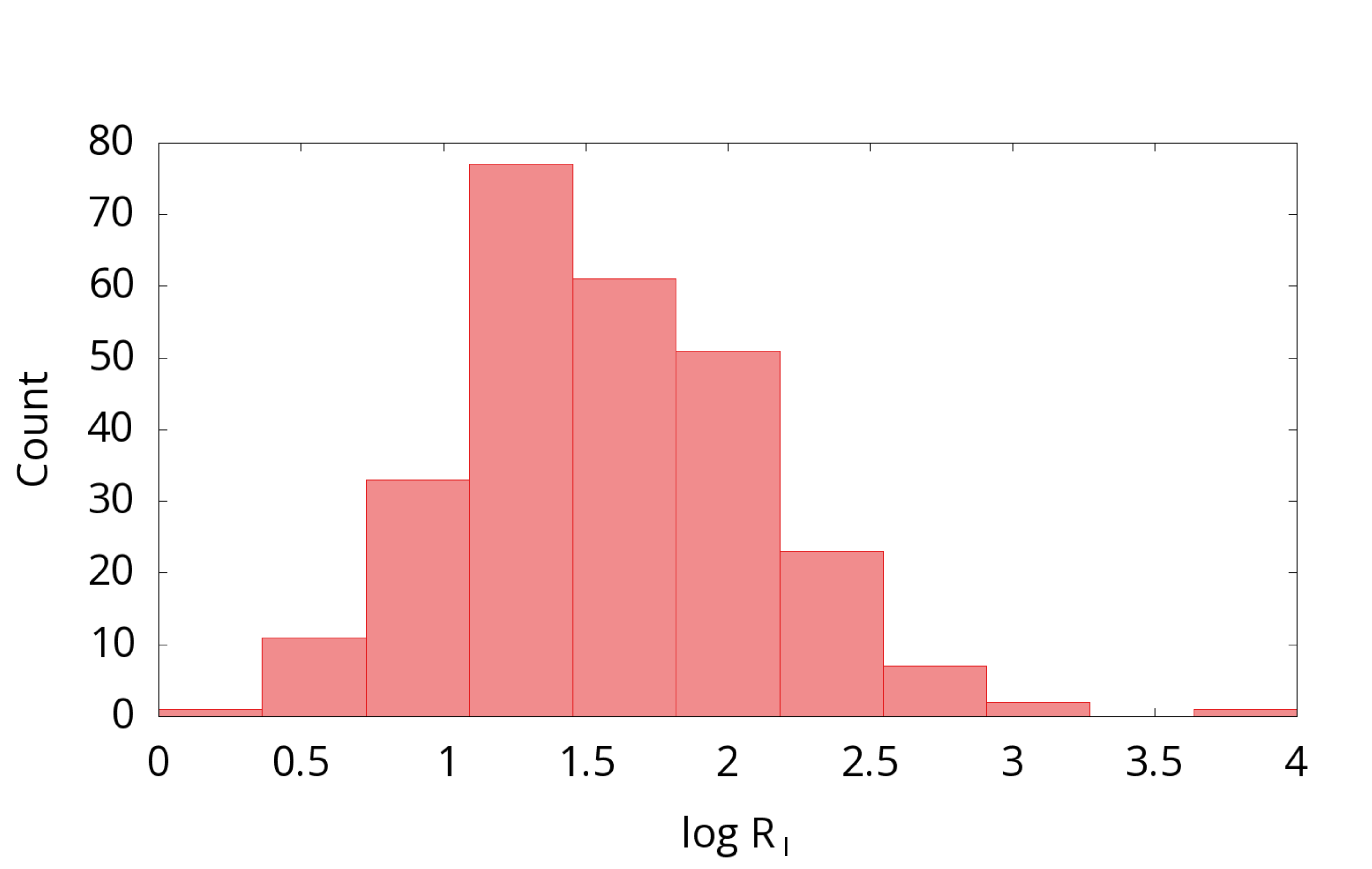}
\includegraphics[width=\columnwidth]{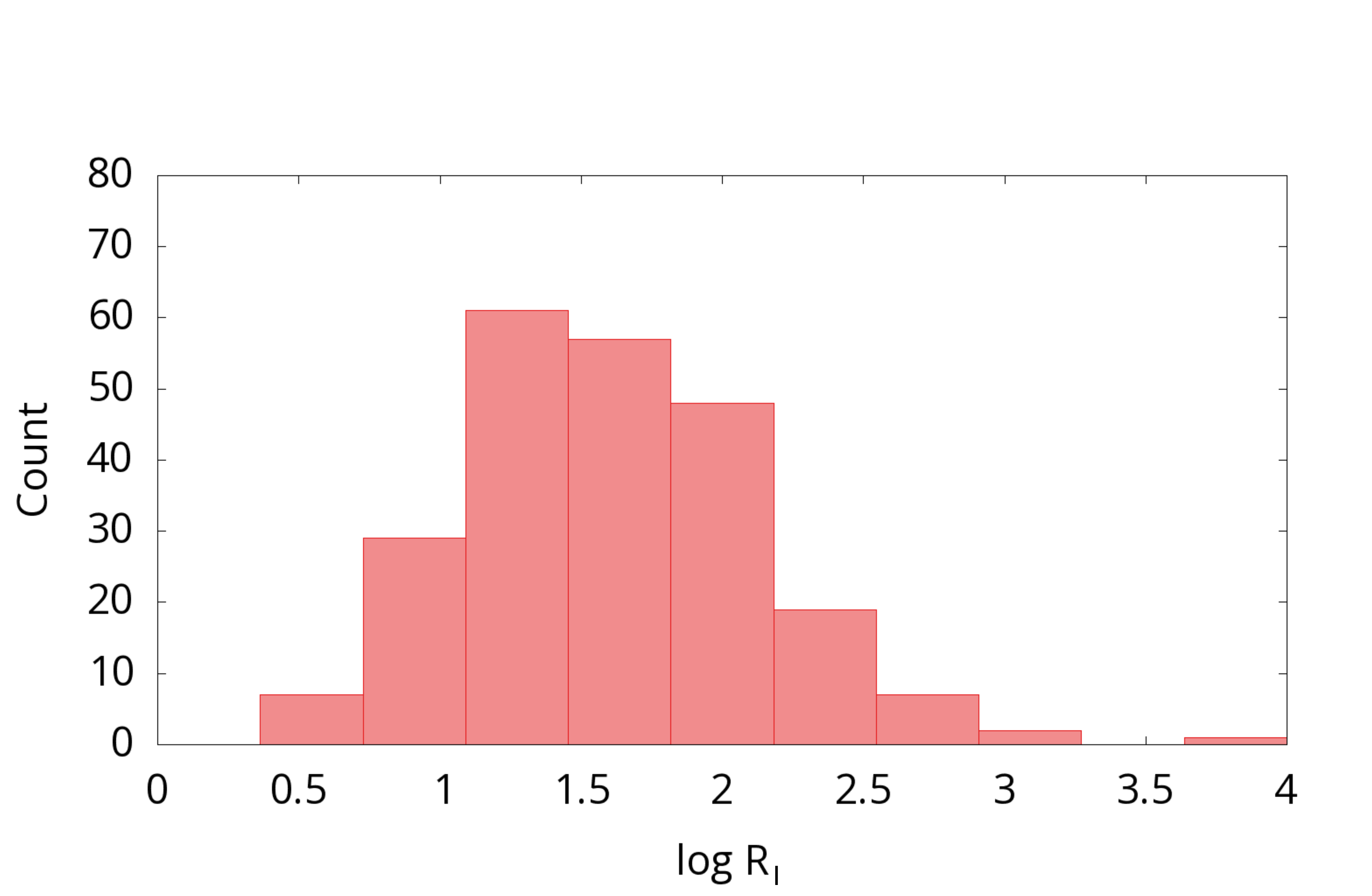}
\includegraphics[width=\columnwidth]{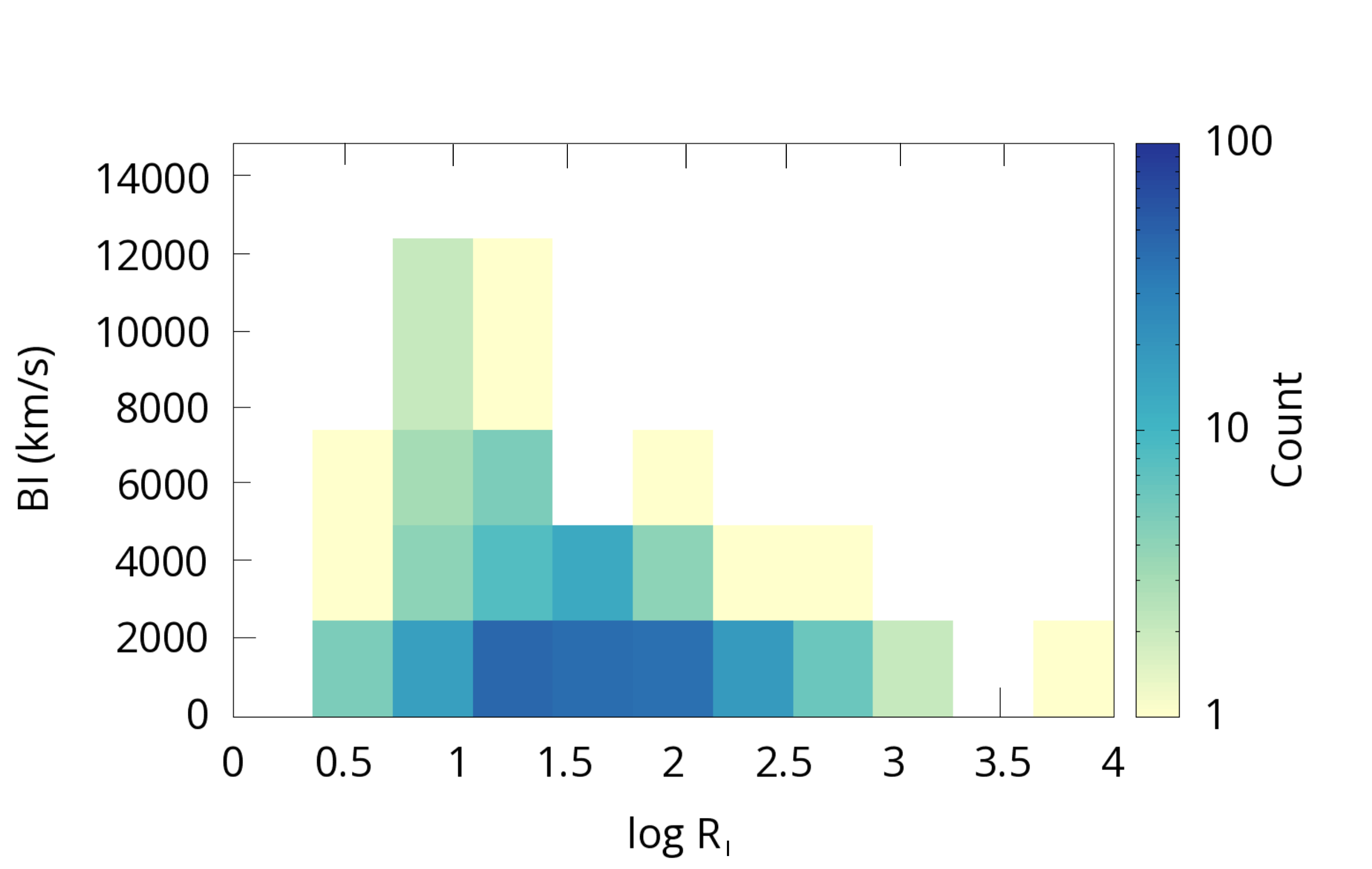}
\caption{Top: Distribution of the radio-loudness parameter $\rm R_{I}$ \citep{kimball2011b} for the whole sample
of radio-loud BALQSOs selected from \citet{gibson2009} as described in Section ~\ref{sample}. Middle: 
the same histogram but limited to the sources with $\rm BI_{0}(C_{IV})>0$. Bottom: 
Radio-loudness vs. the modified balnicity index $\rm BI_{0}(C_{IV})$ for the same 
sample of radio-loud BALQSOs.}
\label{counts}
\end{figure}

Following \citet{kimball2011b} we have calculated the radio-loudness parameter $\rm R_{I}$ defined as the radio-to-optical (i-band) 
ratio of the quasar core for all the BALQSOs selected by us from the most recent catalogue created by \citet{gibson2009}. 
Since all these sources are unresolved on FIRST resolution we assumed that their integrated 1.4\,GHz flux corresponds to a core flux and adopted  the radio core 
spectral index and optical spectral index with  a value of 0 and -0.5, respectively.
The histogram in Fig.~\ref{counts} 
shows the number of BALQSOs versus the radio-loudness parameter $\rm R_{I}$. 
The peak of this distribution is in the range $\rm 1<log\,R_{I}<1.5$. Most of the quasars from the VLBI sample have  values of log\,$\rm R_{I}>2$ (Table~\ref{basic}) so they belong to the tail of log\,$\rm R_{I}$ distribution
for BALQSOs. It has been suggested that the radio-to-optical ratio of the quasar core is a better statistical measure of core boosting and therefore 
of orientation than the previously used core-to-lobe (R) ratio \citep{wills1995}. Recently \citet{kimball2011b} showed that both parameters, $\rm R_{I}$ and R, are correlated which supports the hypothesis that they are indicative of the line-of-sight orientation. Objects with log\,$\rm R_{I}>2.5$ are probably viewed close to the radio jet axis and quasars with log\,$\rm R_{I}<1.5$ are thought to
be those with large viewing angles. The radio-loudness distribution of BALQSOs can indicate that most of them are seen close to the disk plane.  

The core-jet structures of quasars from the VLBI sample corresponds to the higher values
of the radio-loudness parameter, log\,$\rm R_{I}>2$. Our selection criteria caused that they constitute the `most extreme' cases among the BALQSOs. The one exception is the unresolved source 2107-0620.
The low value of log\,$\rm R_{I}$ and its X-ray weakness \citep{wang2008} may imply that this source is viewed close to the plane of the accretion disk. 

\subsubsection{Variability brightness temperatures and viewing angle determination}
The maximum brightness temperature of powerful extragalactic radio sources is limited to the value of $\rm 10^{12}\,K$ by the inverse Compton 
catastrophe. Using the total flux density variations of a sample of extragalactic radio sources \citet{lahteenmaki1999} showed that the individual
values of the intrinsic brightness temperature clustering around the equipartition limit of $\rm 10^{11}\,K$, in accordance with \citet{readhead1994}. 
In the case of Doppler boosted sources ($\rm \delta>1$) the variability brightness temperatures are
much larger than the VLBI brightness temperatures.

We looked for variable radio-loud BALQSOs in the parent sample selected from \cite{gibson2009} using the method described by \citet{zhou2006}. 
We compared the FIRST and the NRAO VLA Sky Survey (NVSS; \citet{condon1998}) fluxes of these quasars in order to find differences
implying the radio variability. The NVSS and FIRST surveys were carried out at the same frequency covering the same sky area but with different flux 
limit of the sources, different resolution, and at different times. We searched for counterparts of the selected quasars in the NVSS within 10'' 
matching  radius.

\begin{landscape}
\begin{table}
{\small
\caption[]{Candidates for variable sources}
\label{variable}
\begin{tabular}{cccccccccccccccccc}
\hline
RA(J2000) & Dec(J2000)&
\multicolumn{1}{c}{\it z}&
\multicolumn{1}{c}{F$_{peak}$}&
\multicolumn{1}{c}{$\sigma_{F_{peak}}$}&
\multicolumn{1}{c}{$\rm Epoch_{F}$}&
\multicolumn{1}{c}{N$_{int}$}&
\multicolumn{1}{c}{$\sigma_{N_{int}}$}&
\multicolumn{1}{c}{$\rm Epoch_{N}$}&
\multicolumn{1}{c}{BI}&
\multicolumn{1}{c}{$\rm BI_{0}$}&
\multicolumn{1}{c}{log $R_{I}$}&
\multicolumn{1}{c}{${\rm T_{b}(var)}$}&
\multicolumn{1}{c}{$\rm \delta_{1}$}&
\multicolumn{1}{c}{$\theta_{1}$}&
\multicolumn{1}{c}{$\rm \delta_{2}$}&
\multicolumn{1}{c}{$\theta_{2}$}\\

h~m~s & $\degr$~$\arcmin$~$\arcsec$ & 
\multicolumn{1}{c}{}&
\multicolumn{1}{c}{(mJy)} &
\multicolumn{1}{c}{(mJy)}&
\multicolumn{1}{c}{} & 
\multicolumn{1}{c}{(mJy)}&
\multicolumn{1}{c}{(mJy)}&
\multicolumn{1}{c}{} & 
\multicolumn{1}{c}{} &
\multicolumn{1}{c}{} &
\multicolumn{1}{c}{}&
\multicolumn{1}{c}{($\rm 10^{11}~\delta^{-3}~K$)}&
\multicolumn{1}{c}{}&
\multicolumn{1}{c}{(deg)}&
\multicolumn{1}{c}{}&
\multicolumn{1}{c}{(deg)}\\

(1)& (2)& (3) &(4)&   
\multicolumn{1}{c}{(5)}&
\multicolumn{1}{c}{(6)}& 
\multicolumn{1}{c}{(7)}&
\multicolumn{1}{c}{(8)}&  
\multicolumn{1}{c}{(9)}&
\multicolumn{1}{c}{(10)}&
\multicolumn{1}{c}{(11)}&
\multicolumn{1}{c}{(12)}&
\multicolumn{1}{c}{(13)}&
\multicolumn{1}{c}{(14)}&
\multicolumn{1}{c}{(15)}&
\multicolumn{1}{c}{(16)}&
\multicolumn{1}{c}{(17)}\\

\hline 
00      40      22.406  &       00      59      39.67   &       2.57    &       9.63    &       0.10    &       1994.684        &       6.4     &       0.5     &       1993.874        &       1       &       1       &       1.86    &       342.9   &       $>$7.00 &       $<$8.2  &       $>$3.25 &       $<$17.9 \\
00      44      44.059  &       00      13      03.55   &       2.29    &       53.08   &       0.11    &       1999.193        &       48.1    &       1.5     &       1993.874        &       1       &       1       &       2.84    &       10.1    &       $>$2.16 &       $<$27.6 &       $>$1.00 &       $-$     \\
07      53      10.422  &       21      02      44.31   &       2.29    &       16.78   &       0.15    &       1998.695        &       14.4    &       0.6     &       1993.836        &       1       &       1       &       1.95    &       5.7     &       $>$1.79 &       $<$34.0 &       $<$1    &       $-$             \\
08      28      17.249  &       37      18      53.64   &       1.35    &       21.18   &       0.13    &       1994.561        &       14.8    &       0.6     &       1993.956        &       0       &       1       &       2.47    &       380.4   &       $>$7.25 &       $<$7.9  &       $>$3.36 &       $<$17.3 \\
08      11      02.931  &       50      07      24.52   &       1.84    &       23.07   &       0.19    &       1997.262        &       19.5    &       0.7     &       1993.874        &       1       &       1       &       2.14    &       12.1    &       $>$2.29 &       $<$25.9 &       $>$1.06 &       $<$70.6 \\
09      05      52.412  &       02      59      31.62   &       1.82    &       43.54   &       0.14    &       1998.542        &       36.4    &       1.2     &       1993.874        &       1       &       1       &       1.48    &       12.4    &       $>$2.32 &       $<$25.5 &       $>$1.08 &       $<$67.8 \\
09      33      48.373  &       31      33      35.30   &       2.60    &       18.35   &       0.14    &       1995.058        &       16.3    &       0.6     &       1993.956        &       1       &       1       &       1.99    &       119.9   &       $>$4.93 &       $<$11.7 &       $>$2.29 &       $<$25.9 \\
09      38      04.534  &       12      00      11.26   &       2.23    &       11.93   &       0.14    &       1999.998        &       10.2    &       0.5     &       1995.159        &       1       &       1       &       1.60    &       4.0     &       $>$1.59 &       $<$39.0 &       $<$1    &    $-$             \\
10      34      24.422  &       49      32      21.13   &       1.62    &       12.42   &       0.13    &       1997.289        &       9.7     &       0.5     &       1993.874        &       0       &       1       &       2.13    &       7.2     &       $>$1.93 &       $<$31.2 &       $<$1    &       $-$             \\
10      43      19.869  &       13      27      49.03   &       1.45    &       7.29    &       0.14    &       1999.976        &       5.7     &       0.5     &       1995.159        &       0       &       1       &       1.36    &       1.7     &       $>$1.20 &       $<$56.4 &       $<$1    &       $-$             \\
10      44      52.417  &       10      40      05.91   &       1.88    &       16.4    &       0.15    &       2000.029        &       13.1    &       0.6     &       1995.159        &       1       &       1       &       1.42    &       5.6     &       $>$1.78 &       $<$34.2 &       $<$1    &       $-$             \\
11      22      20.462  &       31      24      41.19   &       1.45    &       12.64   &       0.13    &       1994.669        &       10.8    &       0.5     &       1993.956        &       1       &       1       &       1.32    &       90.4    &       $>$4.49 &       $<$12.9 &       $>$2.08 &       $<$28.7 \\
13      30      02.512  &       51      03      03.49   &       2.92    &       6.55    &       0.13    &       1997.316        &       4.7     &       0.4     &       1993.874        &       1       &       1       &       1.36    &       13.5    &       $>$2.38 &       $<$24.8 &       $>$1.10 &       $<$65.4 \\
14      01      26.163  &       52      08      34.63   &       2.97    &       36.18   &       0.14    &       1997.337        &       30.4    &       1.0     &       1993.874        &       1       &       1       &       2.02    &       42.8    &       $>$3.50 &       $<$16.6 &       $>$1.62 &       $<$38.1 \\
14      41      36.257  &       63      25      18.76   &       1.78    &       7.69    &       0.23    &       2002.594        &       4.4     &       0.4     &       1993.896        &       0       &       1       &       1.50    &       1.6     &       $>$1.17 &       $<$58.7 &       $<$1    &       $-$             \\
14      44      34.816  &       00      33      05.49   &       2.04    &       12.76   &       0.15    &       1998.567        &       10.5    &       0.5     &       1995.159        &       0       &       1       &       1.90    &       9.1     &       $>$2.09 &       $<$28.6 &       $<$1    &       $-$             \\
14      55      06.755  &       35      47      49.33   &       3.05    &       3.99    &       0.14    &       1994.519        &       2.3     &       0.4     &       1995.290        &       1       &       1       &       1.67    &       263.9   &       $>$6.41 &       $<$9.0  &       $>$2.98 &       $<$19.6 \\
14      59      26.330  &       49      31      36.79   &       2.37    &       5.22    &       0.14    &       1997.291        &       3.8     &       0.4     &       1995.194        &       1       &       1       &       1.24    &       19.6    &       $>$2.69 &       $<$21.8 &       $>$1.25 &       $<$53.1 \\
15      56      33.769  &       35      17      57.64   &       1.50    &       30.9    &       0.15    &       1994.506        &       27.5    &       0.9     &       1995.290        &       1       &       1       &       1.98    &       147.2   &       $>$5.28 &       $<$10.9 &       $>$2.45 &       $<$24.1 \\
17      05      59.013  &       24      35      32.71   &       1.56    &       93.96   &       0.14    &       1995.937        &       85.2    &       2.6     &       1995.162        &       1       &       1       &       2.81    &       417.6   &       $>$7.47 &       $<$7.7  &       $>$3.47 &       $<$16.7 \\
21      07      57.683  &       -06     20      10.49   &       0.65    &       20.29   &       0.15    &       1997.145        &       12.4    &       0.6     &       1993.720        &       1       &       1       &       1.41    &       3.4     &       $>$1.51 &       $<$41.5 &       $<$1    &       $-$             \\

\hline
\end{tabular}
\begin{minipage}{228 mm}
\vspace{0.2cm}
{Description of the columns:
(1) and (2) source coordinates (J2000) extracted from FIRST, (3) redshift as measured from the SDSS, 
(4) FIRST peak flux density, (5) uncertainty of the FIRST peak flux, (6) FIRST observation time, (7) NVSS integrated flux density, (8) uncertainty of the
NVSS integrated flux density, (9) NVSS observation time, (10) ``traditional'' balnicity index selection flag, `1' means BI$>$0 \citep{gibson2009}, (11)
modified balnicity index selection flag, `1' means the value is larger than 0 \citep{gibson2009}, (12) radio-loudness, the radio-to-optical (i-band) ratio of the 
quasar core \citep{kimball2011b}, which were calculated
from {\it z}, ${\rm S_{1.4\,GHz}}$, $\rm M_{i}$ taken from \citet{schneider2007}, and the assumption of a radio core spectral index of 0 and
an optical spectral index of -0.5, (13) lower limit of the brightness temperature calculated using equation \ref{var}, (14) minimum Doppler factor calculated using the equipartition
brightness temperature value of $\rm 10^{11}\,K$, (15) maximum viewing angles estimated as described in the Appendix using the value of $\rm \delta_{1}$, 
(16) minimum Doppler factor calculated using the inverse Compton brightness temperature value of $\rm 10^{12}\,K$, (17) maximum viewing angles estimated as described 
in the Appendix using the value of $\rm \delta_{2}$.
}
\end{minipage} 
}
\end{table} 
\end{landscape}

To classify the radio-loud BALQSO as a variable source we computed the variability ratio (VR) and the significance of the radio flux
variability ($\sigma_{var}$) for 
each quasar as proposed by \citet{zhou2006}. The variability ratio (VR) is defined as 

\begin{equation}
VR=\frac{F_{peak}}{N_{int}}
,\end{equation}

where $F_{peak}$ is the FIRST peak flux density and $N_{int}$ is the NVSS integrated flux density, and the significance of the radio flux variability ($\sigma_{var}$) is

\begin{equation}
\sigma_{var}=\frac{F_{peak}-N_{int}}{\sqrt{{\sigma_{F_{peak}}^2}+{\sigma_{N_{int}}^2}}}
,\end{equation}

where the $\sigma_{F_{peak}}$ and $\sigma_{N_{int}}$ are the uncertainties of the FIRST peak flux densities and NVSS integrated flux densities, respectively.
BALQSOs with $\rm VR>1$ and $\sigma_{var}>3$ were selected as candidates for variable sources. Finally, the sample consists of 21 candidates for variable BALQSOs and
is presented in Table~\ref{variable}.

Using equation \ref{var} (Appendix) we have calculated variability brightness temperature for all these objects. We then estimated the viewing angle $\theta$ of each 
object, defined as an angle between the jet axis and the observer,
as the maximum of the  function \citep{ghosh2007}

\begin{equation}
cos\theta \leqslant \max\left[ \frac{\delta - \sqrt{1-\beta^2}}{\delta \beta}\right]
\label{theta}
,\end{equation}

where $\beta$ is the velocity of the jet and $\delta$ is the \textcolor{red}{ }Doppler factor. As was discussed at the beginning of this section, the intrinsic 
brightness temperature of extragalactic radio sources is probably in the range of $\rm 10^{11} - 10^{12}\,K$ and this allows us to estimate the minimum 
Doppler factor that avoids the inverse Compton catastrophe to be $\rm \delta_{1}=\Big(\frac{T_{b}(var)}{10^{11}\,K}\Big)^{1/3}$ and 
$\rm \delta_{2}=\Big(\frac{T_{b}(var)}{10^{12}\,K}\Big)^{1/3}$, respectively. 
We then looked for the maximum value of function \ref{theta} for both $\rm \delta_{1}$ and $\rm \delta_{2}$ in order to 
determine the range where the maximum viewing angle of the quasar should be found (see  Appendix for details).
 
\begin{figure}
\centering
\includegraphics[width=\columnwidth]{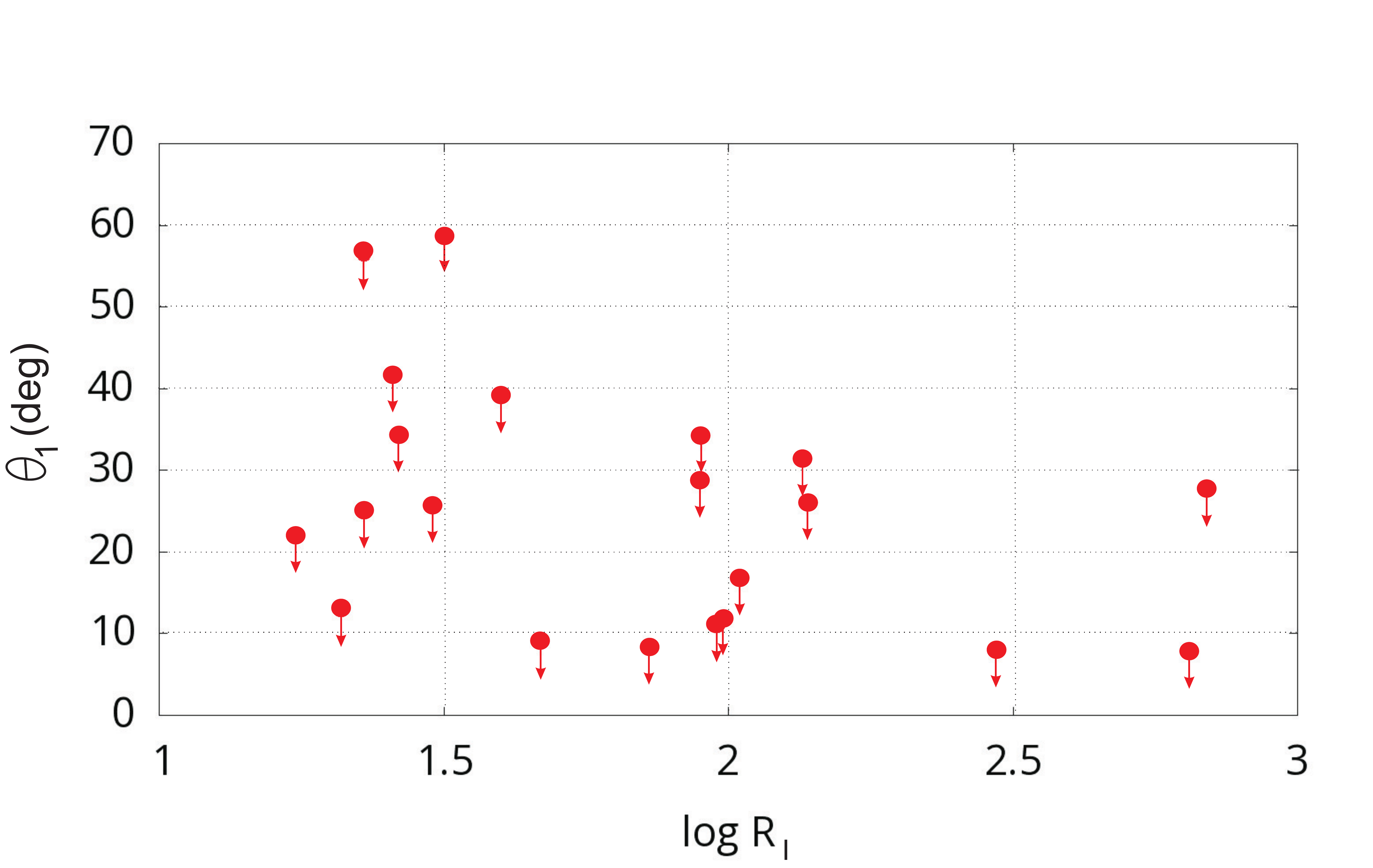}
\caption{Upper limits of the viewing angles ($\theta_{1}$) vs. radio-loudness parameter $\rm R_{I}$ for the sources from Table~\ref{variable}. }
\label{range}
\end{figure}

As a result we have obtained a wide range of viewing angles reaching the value $\rm \ga 45\degr$ in a few cases. 
The most populated subgroup of BALQSOs with log\,$\rm R_{I}<1.5$ have the maximum viewing angle in the range $59 - 68\degr$ which constitutes the lower limit.
This means that the viewing angle of the radio weakest BALQSOs is in the range $0 - 68\degr$ (see also Fig.~\ref{range}).

Three of the BALQSOs from the VLBI sample, namely the 0811+5007, 1401+5208, and 2107-0620, also show flux variability and are listed in Table~\ref{variable}. A wide range of viewing angles is estimated in their case.
We have already mentioned that the lack of X-ray detection of 2107-0620 can be caused by the presence of the shielding gas \citep{wang2008}. There are no X-ray observations of 1401+5007, but the 0811+5007 has been detected 
in X-rays showing no clear evidence for X-ray absorption  from neutral hydrogen. As an explanation, \citet{wang2008} suggested that if absorption is present 
in this source it must be more complex or that the X-rays come from the radio jet outside the shielding gas. 
Interestingly, the broad absorption line disappearance has  recently been reported in 0811+5007 \citep{filiz2012}. Such variations of the BAL troughs could be caused by the 
disk-wind rotation or variations of shielding gas that lead to variations of ionizing-continuum radiation.
However, the second explanation cannot be ruled out. As has  already been discussed in the case of 
BALQSO 1045+352 \citep{kunert2009}, the jet synchrotron self-Compton (SSC) emission can be
quite significant in some quasars and may dominate the X-ray energy range, while the X-ray emission from the corona is absorbed in a large part 
(see also \citet{miller2009}).
The comprehensive study of 1045+352 also revealed  the re-start of the jet activity in this source with the viewing angle $\sim 30^{\rm o}$ \citep{kun2010a}.

As expected from previous studies \citep{shankar2008, ghosh2007} the number of candidates for variable BALQSOs is small, but even this small number can be overestimated \citep{ofek2011}. Although the methodology proposed by \citet{zhou2006} is a good start in looking for variability of radio sources, 
only a long-term monitoring programme of flux density of candidate sources from Table~\ref{variable} will be able to provide more accurate measurements of the variability time scale, value of the flux change, etc., and thereby the $\theta_{max}$ value. Further observational evidence for the existence of boosted radio emission in some BALQSOs has been found \citep{reynolds2013,berrington}.

\subsection{Discussion}
\label{balnicity}

\citet{gregg2006} has noted the scarcity of broad absorption lines (BALs) among the large-scale, lobe-dominated radio-loud quasars and a drop in the BI index
with increasing radio-loudness of the sources. Based on these results they proposed an evolutionary scenario in which BALs are associated with an early
stage of the quasar evolution when a young quasar emerges from the surrounding dust. The anticorrelation between BI index and radio-loudness has not 
been confirmed by the statistical studies of \citet{shankar2008} based on the \citet{trump2006} catalogue. However, as has been pointed out by 
\citet{gibson2009} there is a paucity of quasars that are simultaneously strongly radio-loud and heavily absorbed.
The bottom panel of Fig.~\ref{counts} shows the distribution of balnicity index versus the radio-loudness parameter log\,$\rm R_{I}$ for quasars selected 
by us from \citet{gibson2009}.
It can be seen that the largest values of BI index occurs among objects with log\,$\rm R_{I}<1.5$.
The radio variability study of a sample of BALQSOs indicates that the same subgroup of BALQSOs has the widest range of viewing angles $\theta$ (Fig.~\ref{range}). This range changes as a function of the radio-loudness parameter for the whole sample, which seems to be in agreement with recent results of the radio-loudness parameter as an orientation indicator \citep{kimball2011b}.
It is commonly believed that the division line on quasars and radio galaxies should be at $\theta=40-45\degr$ \citep{barthel1989,ghisellini} and our limits ($<59 - 68\degr$) are consistent with that.  
Earlier Monte Carlo simulations of viewing angles of BALQSOs showed generally smaller values for both BAL and non-BALQSOs ($<33\degr$), but pointed out that the viewing angles 
of BALQSOs may extend $\sim \rm 10\degr$ farther  from the radio jet axis than non-BAL objects \citep{dipompeo2012}. Recent numerical simulations assuming a simple kinematic disk wind model can predict BAL spectra for a variety of viewing angles, as large as $\sim \rm 80\degr$ \citep{higgin}.

We would like, however, to draw attention to the fact that the most numerous log\,$\rm R_{I}<1.5$ subgroup of radio-loud BALQSOs is not exclusively connected with strong absorption. 
In practice, we observe a very wide range of BI indices among them and we conclude that orientation is only one of the factors influencing the measured absorption. 
In other words, the largest viewing angles that still guarantee
the source to be classified as a quasar do not guarantee the largest absorption in every case. Spectroscopic and variability studies of BAL features shows
that complex behaviour of wind outflows, from minor depth changes within the BAL profile to BAL disappearance \citep{capellupo2013, filiz2012}, is present in BALQSOs and
can be responsible for a wide range of covering factors within objects from the same group characterized by e.g. the radio-loudness parameter $\rm R_{I}$.

The selection criteria of our sample ($\rm S_{FIRST}>2\,mJy$),  based on the \citet{gibson2009} catalogue, led to only $\sim\,15\%$ of the sources being formally
radio-quiet ($\rm log\,R<1$). In fact, this number is much larger \citep{welling2014}, although deep radio observations and population analysis show that all
quasars have some minimum level of radio flux \citep{White2007,kimball2011a,condon2013}. The origin of the radio emission in sources weaker than 2.4\,mJy
could be  stellar \citep{condon2013},  while in the case of stronger objects -- with $\rm logL_{1.4\,GHz}>23.4\,W\,Hz^{-1}$ -- 
the radio activity is powered  not by starbursts \citep{kimball2011a,ulvetsad} but  by AGNs or may originate in shocks formed by the collision of the disk wind and 
the interstellar medium \citep{sikora2009,zakamska2014}. The radio properties of the luminous BALQSOs presented in this paper indicate that  their radio
emission is of AGN origin rather than  stellar. However, when considering our whole  sample it can be seen that many of the radio-loud BALQSOs are intermediate-radio objects 
with 1.4\,GHz luminosities in the FR\,I - FR\,II transition region ($\rm 24<logL\,[W\,Hz^{-1}]<25.5$). 
Some of them, with flat spectrum, could in  fact be  boosted weak sources \citep{falcke1996, berrington}, but the majority have steeper spectra than  
non-BALQSOs \citep{dipompeo2011, bruni2012}. What is more, they are probably young objects at the beginning of their evolution.  

Studies of the compact radio sources suggest that their evolution is determined by the properties of their central engine: strength, accretion mode,
excitation level of the ionized gas, and the ISM \citep{kun2010b,Kunert10b}.  Low luminosity young AGNs may develop  diffuse, large-scale structures 
\citep{gawron2006,cegla2013} as their weak jets are disrupted before escaping their host galaxies. They may undergo disrupted evolution many times, before finally they will be able to escape 
the host galaxy and evolve further. Such behaviour explains the excess of compact AGNs compared to large-scale FR\,Is and FR\,IIs \citep{odea98,kun2010b}. According to \citet{shankar2008} there is no difference between the 
fraction of large-scale sources among the BALQSOs and among the overall sample of radio objects at fixed 
radio power. Therefore, we suggest that the short lifetime of some compact AGNs could also explain the low number 
of extended `adult' radio sources among the BALQSOs. In this sense some of the compact AGNs (and hence BALQSOs) can be also considered as the 
re-activated objects as proposed by a few authors \citep{kun2010a, bruni2013, hayashi2013} based on the larger-scale extended emission found around the new compact source.

\begin{figure}
\centering
\includegraphics[width=\columnwidth]{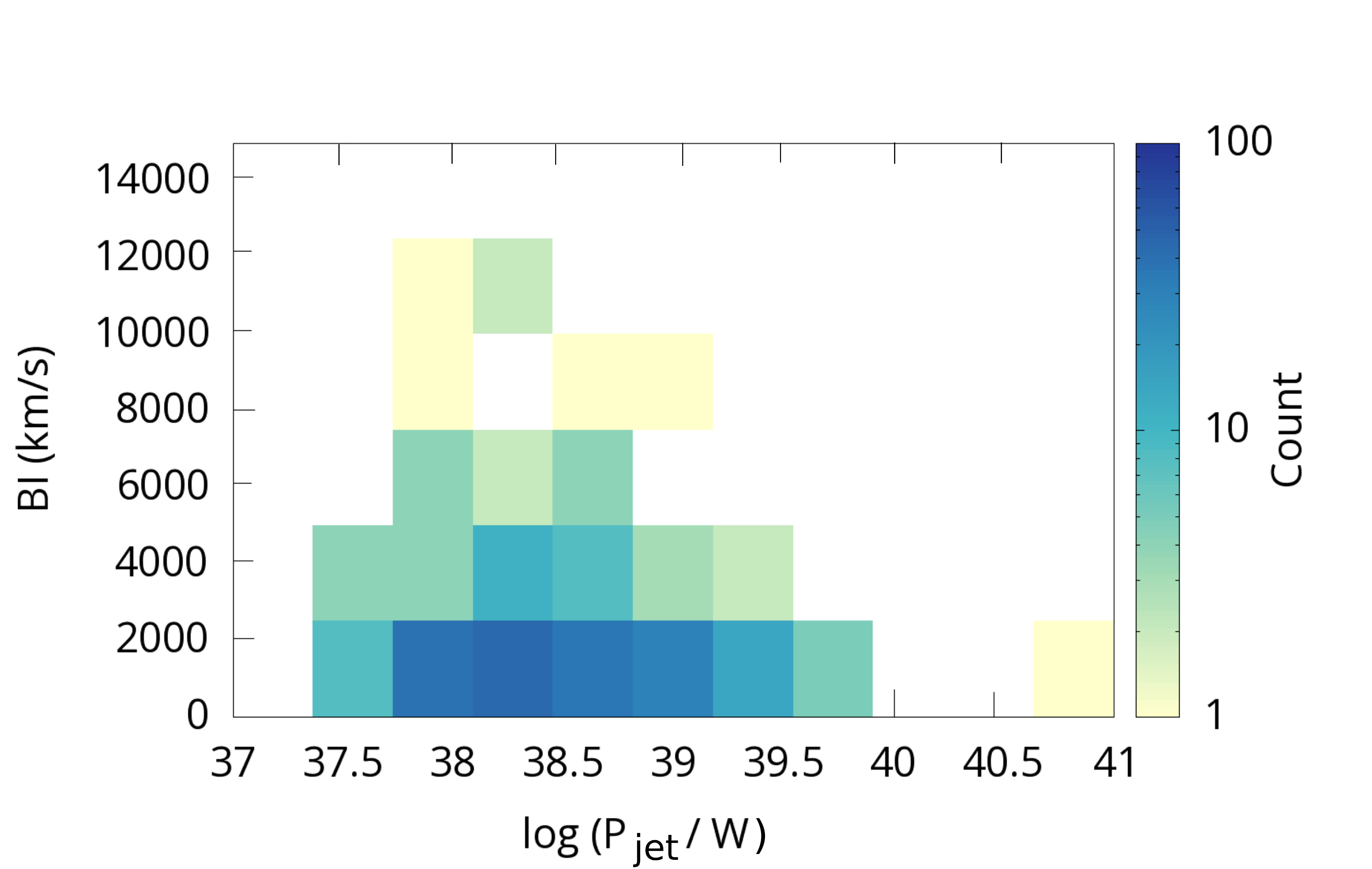}
\caption{
Jet power vs. the value of modified balnicity index $\rm BI_{0}(C_{IV})$ for the whole sample
of radio-loud BALQSOs selected from \citet{gibson2009} as described in Section ~\ref{sample}. The jet power is calculated for f=10 \citep{sikora2013}.}
\label{jet}
\end{figure}

According to the analysis performed by \citep{willott1999}, the radio luminosity of a large-scale radio source is approximately proportional to the jet power. Adopting the 
modifications made by \cite{sikora2013}, we use their equation $\rm P_{jet}\sim 10^{2}\,\nu_{1.4\,GHz}\,L_{1.4\,GHz}\,(f/3)^{3/2}$--  where factor {\it f} is in the range 1- 20 --
to calculate the jet power of the selected BALQSOs. We then plot the 1.4\,GHz jet power versus the value of modified balnicity index $\rm BI_{0}(C_{IV})$ and present it
in  Fig.~\ref{jet}. There is no  formal correlation between these two parameters, but the trend of the highest values of BI index to be associated with the lower jet powers
can be seen in accordance with the scenario discussed above.\footnote {
The Willott formula is only accurate if there are relaxed radio lobes. Since most of our sources have sub-galactic sizes, the possible jet-dense medium interaction can increase the radio luminosity and thus the calculated jet powers can be overestimated. This, however, should  affect the whole group equally and will not change the visible trend.}

\section{Summary}
We have presented EVN and VLBA observations of a sample of radio-loud BALQSOs together with the variability studies of some of them. 
We find the following:

$\bullet$
The peak of the radio-loundness parameter (log\,$\rm R_{I}$) distribution of our sample of compact BALQSOs is in the range 1 - 1.5. The number of sources 
decreases with the increasing radio flux density and log\,$\rm R_{I}$. Therefore, the BALQSOs that we are able to image with the VLBI are the most luminous ones,
which constitute the minority of the radio-loud BALQSO population.

$\bullet$
Nine of the ten sources from the VLBI sample are strong radio-loud BALQSOs with 1.4\,GHz
flux densities in the range 20 - 80\,mJy and log\,$\rm R_{I}>1.8$. The one exception is the low redshift source 2107-0620. All sources have very compact sizes ($<$50\,pc), but most of them have been resolved
at 5\,GHz showing one-sided, probably core-jet structures, typical of quasars. 

$\bullet$
Distribution of the BI index versus the radio-loudness parameter, log\,$\rm R_{I}$, shows that the strongest absorption is associated with the lower values of the radio-loudness parameter, 
log\,$\rm R_{I}<1.5$. The $\rm R_{I}$ parameter is considered  an indicator of orientation and if this is the case the values of log\,$\rm R_{I}<1.5$ can mean large viewing angles.   
However, the large span of BI values in the each bin of the radio-loudness parameter indicates that orientation is only one of the factors that influence the measured absorption.

$\bullet$
Many of the radio-loud BALQSOs are intermediate- or low-power radio objects with 1.4\,GHz luminosities in the FR\,I - FR\,II transition region and below it. They are powered by 
the AGN and their compact sizes may imply young age. Studies of the low luminosity compact radio sources indicate that many of them can be short-lived objects 
that undergo the GPS/CSS phase of activity many times before they become large-scale FR\,I or FR\,II sources. Such intermittent behaviour may also account  for the 
scarcity of extended radio sources among the BALQSOs.

$\bullet$
The fact that the broad absorption lines are present mostly in the low-power radio objects may suggest an anticorrelation between the power of the jet and the value of the absorption and an indirect correlation  with the process responsible for BAL phenomenon. We do not find such a relationship in our data; however, 
there is a hint in our analysis that the highest values of BI index are associated with the lower jet powers in BALQSOs.

\section{Acknowledgments}
This work was supported by the National Scientific Centre under grant DEC-2011/01/D/ST9/00378.\\
The research leading to these results has received funding from the European Commission Seventh Framework Programme (FP/2007-2013) under grant agreement No 283393 (RadioNet3).\\
The European VLBI Network is a joint facility of European, Chinese, 
South African, and other radio astronomy institutes funded by their national research councils.\\
The National Radio Astronomy Observatory is a facility of the National Science Foundation 
operated under cooperative agreement by Associated Universities, Inc.\\
The research described in this paper makes use of Filtergraph, an online data 
visualization tool developed at Vanderbilt University through the Vanderbilt 
Initiative in Data-intensive Astrophysics (VIDA).\\

\appendix
\section{Brightness temperature and viewing angle}

\subsection{Brightness temperature}

The observed flux density of a spherical or cylindrical source with 
radius $r$ is proportional to the source surface and inverse proportional
to the square of the luminosity distance ($D_{\rm L}$)
\begin{equation} 
F_{\nu} = \frac{\pi r^2}{D^2_{\rm L}} I_{\nu},
\end{equation}
where $I_{\nu}$ is the intensity of the emission. For a source that moves
with relativistic velocity ($v \la c$) at extragalactic distance,
we have to transform this parameter from the source comoving frame (primed 
quantities) to the observer's frame $I_\nu = (1+z) \delta^3 I'_{\nu'}$,  
where $\delta$ is the Doppler factor and the frequency transformation has
the form 
\begin{equation}
\nu' = \frac{(1+z)}{\delta}\nu.
\end{equation}
In the Rayleigh-Jeans limit ($h\nu \ll kT$) the intensity of the synchrotron
emission can be approximated by a simple relation $I'_{\nu'} = 2 k_{\rm B} \nu' T_b / c^2 $,
where $T_b$ is the brightness temperature, which at radio frequencies should be
comparable with the electron temperature $T_{\rm e}$, where the electron energy
is given by $E = \gamma m_{\rm e} c^2 = 3 k_{\rm B} T_{\rm e}$. Substituting the above relations
in the initial equation we obtain
\begin{equation}
F_{\nu} = \frac{2 k_{\rm B}}{c^2} \frac{\pi r^2}{D^2_{\rm L}} (1+z)^3 \delta \nu^2 T_b.
\label{app_eq3}
\end{equation}
It is useful to use the observed angular size of a source ($\theta$) instead of
the luminosity distance $D_{\rm L} = (1+z)^2 D_{\rm A}$, where $D_{\rm A}$ is
the angular distance. The angular distance can be approximated by
$D_{\rm A} \simeq 2r/\theta$, which finally gives
\begin{equation}
F_\nu = \frac{k_{\rm B} \pi}{2c^2} \frac{\delta}{1+z} \theta^2 \nu^2 T_b.
\label{vlbi}
\end{equation}
If the observed flux density ($F_\nu$) exceeds value of the flux density
($F^{12}_\nu$) obtained for the maximum possible temperature ($T_b \simeq 10^{12} 
{\rm K}$), then we can estimate a minimum value of the Doppler factor
\begin{equation}
\delta \geqslant \frac{F_\nu}{F^{12}_\nu},
\end{equation}  
assuming that the observed excess comes from the beaming effect.

\subsection{Variability time scale}

An upper limit for the source size may be independently estimated from
observed variability time scales. Time of a flux density variation    
($\Delta t'$) in the comoving frame cannot be faster than the source  
crossing time ($r/c$). Therefore, using the time transformation 
($t=(1+z) \Delta t'/\delta$) we can write
\begin{equation}
r \leqslant \frac{\delta}{1+z} c \Delta t,
\end{equation}
where $\Delta t$ is the variability time scale in the observer's
frame. Adopting the previously derived Eq. \ref{app_eq3} we can write
another constraint for the observed flux,
\begin{equation}
F_\nu \leqslant 2 k_{\rm B} \pi \frac{\delta^3}{(1+z)^3} \frac{1}{D_{\rm A}} \Delta t^2 \nu^2 T_b.
\label{var}
\end{equation}

If the observed flux exceeds the value calculated from this relation for
$T_b= 10^{12} \rm K$ or $ T_b= 10^{11} \rm K$ \citep{lahteenmaki1999}, then very likely the emission is beamed and again
we can obtain a lower limit for the Doppler factor.

\begin{figure}
\centering
\includegraphics[width=\columnwidth]{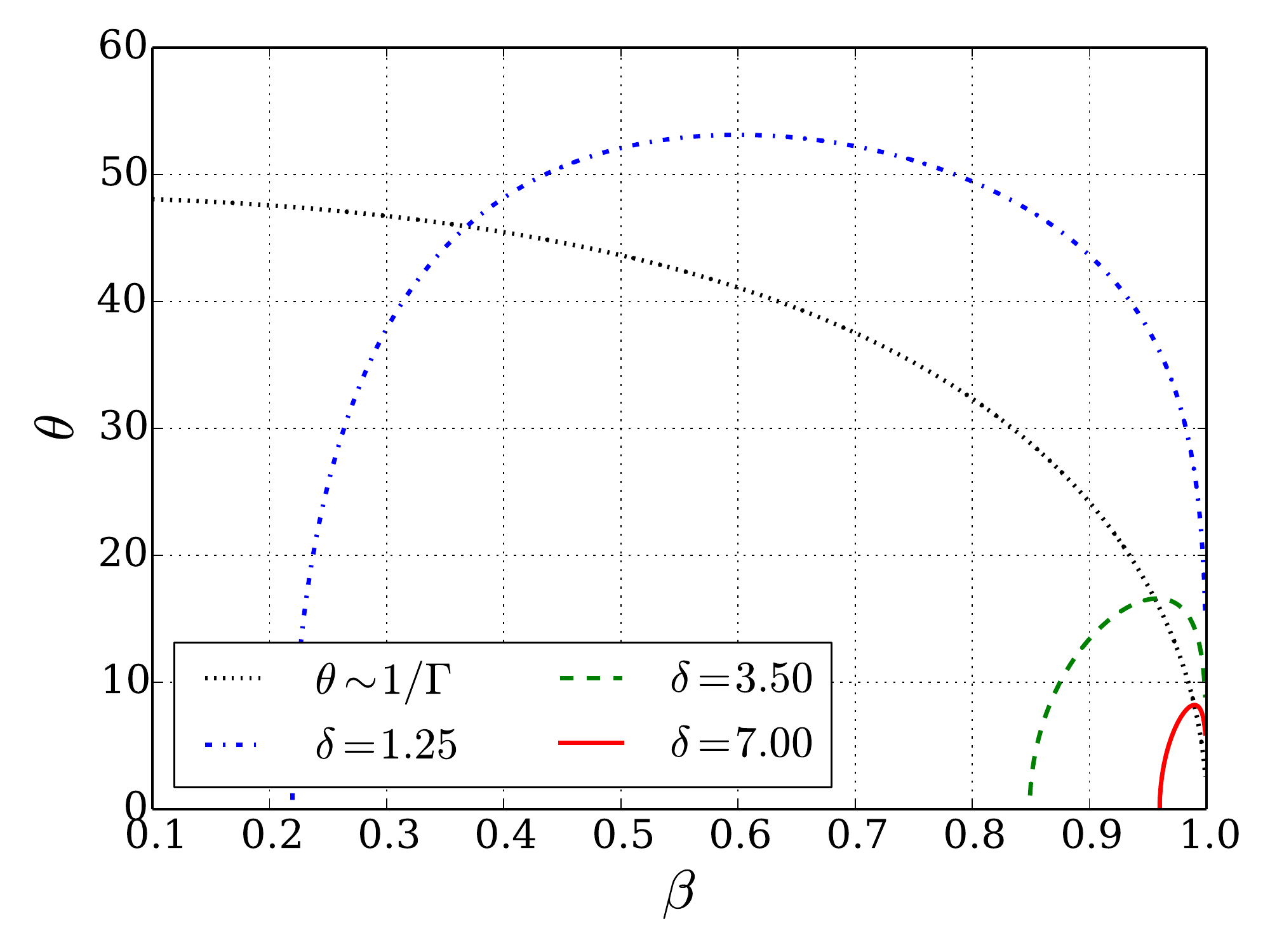}
\caption{Viewing angle $\rm \theta$ versus the jet velocity $\rm \beta$ calculated from equation \ref{max}. }
\label{theta_picture}
\end{figure}

\subsection{Viewing angle}

The Doppler factor depends on the source velocity ($\beta = v/c$) and the
viewing angle ($\theta$). Therefore, an independent estimation of this   
parameter does not provide a direct value of the viewing angle. For a given
value of the Doppler factor we can derive only an upper limit for the viewing
angle
\begin{equation}
cos\theta \leqslant \max\left[ \frac{\delta - \sqrt{1-\beta^2}}{\delta \beta}\right].
\label{max}
\end{equation}

For example, the Doppler factor value $\delta \geqslant 2$  limits
$\theta$ to the $0 - 30^\circ$ range. We note that the viewing angle cannot be
larger than the half of the emission beam angle $\phi \sim 1/\Gamma = \sqrt{1-\beta^2}
\rm \; [rad]$. However, this condition is valid only for $\delta \ga 2$ and
does not give additional constraints to the $\theta$.

\label{lastpage}

\end{document}